\documentclass[amssymb,12pt]{article}

\usepackage{amssymb}

\begin{document}

\title{\textbf\large 
\vspace*{-0.7cm}
\hspace*{12.9cm}{\it\small IAE-6359/2}\\
\vspace*{0.9cm}
Mass spectra of excited 
meson states consisting of $u-,$ $d- $quarks and antiquarks}

\author{ V.V.~Khruschov$^{a,b,}$
\thanks{e-mail: khru@imp.kiae.ru}}
\date{$^{a}$ \textit{\small Russian Research Centre  "Kurchatov
Institute", Kurchatov Sq. 1, Moscow, 123182, Russia}\\
\parindent=-3.4cm $^b$ \textit{\small Center for Gravitation and Fundamental 
Metrology VNIIMS,
Moscow}}

\maketitle
\vspace{-0.5cm}
\begin{abstract}
Mass spectra of excited meson states consisting of $u-$, $d- $quarks and
antiquarks with $I=1$ are considered. The comparison between the existing
experimental data and the mass values evaluated with phenomenological
formulae is carried out. Problems of identification of vector and
scalar meson excited states are discussed.
\end{abstract}

\vspace{1.1cm}

\textbf{1. Introduction} \smallskip

QCD results concerning perturbative calculations of characteristics for
different processes at high energies are widely used, however, considerable 
difficulties exist in the QCD framework, for example,
in hadron spectroscopy  for  mass evaluations  of hadrons consisting of light
quarks. Mainly these difficulties are originated from the unsolved
confinement and spontaneous chiral symmetry breaking problems. Nevertheless,
a number of relations among hadron masses have been obtained using symmetry
or phenomenological considerations. 
For instance, let us remind the mass relations \cite{r1} for Regge
trajectories \cite{r2} or the Gell-Mann-Okubo relation \cite{g1,g2}.
According to the Regge trajectories approach a hadron with its spin $J$ and
mass $M$ within some errors belong to a straight trajectory on the $%
(J,M^{2})-$plane with a slope $\alpha ^{\prime }$ and a intercept $\alpha
_{0}$ 
\begin{equation}
J=\alpha _{0}+\alpha ^{\prime }M^{2}  \label{regge}
\end{equation}
Some hadrons 
belong to trajectories, so called, daughter trajectories, which are roughly
parallel to the main trajectory and are distinguished with different values
of a radial quantum number $n^{r}$ or $n$, $n=n^{r}+1$ \cite{r3,r4,r5}.

Now there is a growing interest in improving the accuracy of existing mass
relations and obtaining these relations in the framework of the QCD or QCD
inspired models. For instance, on the base of QCD finite sum rules \cite%
{k0,k1} the following mass formulae for radially excited $\rho -$ and $\pi -$%
mesons have been obtained

\begin{eqnarray}
M^{2}(\rho ^{n}) &=&M^{2}(\rho )(2n^{r}+1),  \label{kataev} \\
\qquad M^{2}(\pi ^{n}) &=&M^{2}(\pi ^{\prime })n^{r},
\end{eqnarray}%
where $M^{2}(\rho )$ is the mass squared of the $\rho -$meson, $M^{2}(\pi
^{\prime })$ is the mass squared of the first radial excitation of $\pi -$%
meson. In the work \cite{k2} more precise formulae for radially excited
hadrons have been proposed. These formulae in general case can be written as
\begin{equation}
M_{n}^{2}=M_{0}^{2}+\mu ^{2}(n-1)  \label{anisovich}
\end{equation}%

\noindent and describe trajectories on the $(n,M^{2})-$plane with different $M_{0}$
values and approximately the same $\mu $ for each trajectory similarly
Chew-Frautschi plots (\ref{regge}) on the $(J,M^{2})-$plane.

In the present paper we evaluate mass spectra of excited meson states
 consisting
of $u-$, $d-$quarks and antiquarks with mass formulae, which have the
structure peculiar for mass formulae  of the independent
quark model \cite{c1}. These formulae have been proposed for the first time
in the papers \cite{c2} and represent meson mass values in the $(L,M,n)-$%
space. We improve the accuracy of calculations of masses taking more 
precise values
for formulae parameters, compare with existing data and make predictions for
mass values of  meson states, which have no experimental evidence.
Besides that, we consider a possible  compound structures
of some mesons, which are under debates at present, such as scalar mesons and
vector exited mesons.

\medskip
\textbf{2. Phenomenological mass formulae for superlight }$\overline{q}%
q^{\prime }-$\textbf{mesons \smallskip }

In the standard quark model the ground state and exited mesons are formed
from a quark and an antiquark and can be characterized with values of
internal radial and orbital quantum numbers $n^{r}$ and $L$. The total spin $%
S$ of the pair $\overline{q}q^{\prime }$ may be $1$ or $0$. The space and
charge parities of mesons can be evaluated with formulae $P=(-1)^{L+1}$, $%
C=(-1)^{L+S}$ and any $\overline{q}q^{\prime }-$meson with quantum numbers $%
J^{PC}$ may be classified as $n^{2S+1}L_{J}-$state, where $J$ is a meson
spin value. Only superlight mesons, which are made from $u-$, $d-$quarks and
antiquarks, are considered below. Isotopical spin values for mesons of such
kind are $1$ or $0$. We restrict ourselves to those mesons, which have
isotopical spin values equal to $1$, in order to bypass problems \
concerning unknown in some cases mixing parameters for different mesons with
the zero isotopical spin. Further on we neglect mass splittings within
isotopic multiplets, so systematical errors of the order of $10$ $MeV$ for
the phenomenological scheme considered are without doubt admissible.
Moreover, take into account the existing experimental errors for meson
masses, we assume $30$ $\div $ $40MeV$\ as the typical absolute errors of
the mass values evaluations.

In the framework of relativistic independent quark model 
a mass formula for $\overline{q}q^{\prime }-$me\-sons has a structure, which is
different from  structures of  mass formulae in other types of
potential models. For instance, in Ref. \cite{c2} the following mass formula
has been proposed
\begin{equation}
M(n^{2S+1}L_{J})=E_{1}(n_{1}^{r},j_{1},c,\varkappa
)+E_{2}(n_{2}^{r},j_{2},c,\varkappa ),  \label{mnf}
\end{equation}%
where the mass terms (or energy spectral functions) $E_{i}(n_{i},j_{i},c,%
\varkappa ),i=1,2,$ for a quark and an antiquark are defined as

\begin{equation}
E_{i}(n_{i},j_{i},c,\varkappa )=\left\{ 
\begin{array}{c}
c+\varkappa \sqrt{2n^{r}+L+j-1/2},\quad L+j-1/2=2k,\quad k=0,1,... 
\\ 
\varkappa \sqrt{2n_{r}+L+j-1/2},\quad L+j-1/2=2k+1,\quad k=0,1,...%
\end{array}%
\right. 
\label{enf}
\end{equation}

Functions $E_{i}(n_{i}^{r},j_{i},c,\varkappa)$ , $i=1,2$, give the relativistic
effective energies of the quark and antiquark moving in the mean field
inside the meson, and include also an energy of the mean field and possible
nonpotential corrections, which cannot be evaluated within the mean field
approximation. Possibly, a part of these corections can be taken into account
with the help of constant $c$. Note that nonpotential corrections are the most 
important for mass spectra evaluations of light mesons \cite{fil,diek}.
 $n_{i}^{r}$ and $j_{i}$ are a radial quantum number
and a quantum number of an angular moment of the $i-$th constituent,
correspondingly, and $n_{1}^{r}=n_{2}^{r}=n-1$ for the $n{}^{2S+1}L_{J}$ -
meson state, $\varkappa$  is some constant. In general case the root part of 
mass term  $E_{i}(n_{i}^{r},j_{i},c,\varkappa)$, which represent the energy of 
the $i-$th constituent in the mean field,
should be determined from the solution of the Dirac equation \cite{c3}.
However, for the superlight mesons the phenomenological energy spectral
functions $E_{i}(n_{i},j_{i},c,\varkappa )$ in the form (5) are suitable
with a rather good accuracy.

In order to exclude superfluous meson states, the following selection rules
for \textbf{\ }$\overline{q}q^{\prime }-$me\-sons with given $J^{PC}$ values,
quark masses $m_{1}$ and $m_{2}$ and quantum numbers $j_{1}$ and $j_{2}$ are
used
\begin{equation}
\begin{array}{l}
j_{1}=j_{2}=J+1/2,\quad if\quad J=L+S, \\ 
j_{1}=j_{2}+1=J+3/2,\quad if\quad J\neq L+S,\quad m_{1}\leq m_{2}%
\end{array}
\label{jnf}
\end{equation}

One can obtain with the help of formulae (\ref{mnf}), (\ref{enf}) and (\ref%
{jnf}) a few mass relations, which are fulfilled within the systematical
errors of the phenomenological scheme considered. For instance, in the case
of orbitally exitated vector mesons formulae (5) and (6) give the $\rho -$%
trajectory, for radially exitated vector mesons they bring\ to the formula
(2). However, in general case in the framework of this approach instead of
the meson trajectories in $(J,M^{2})-$ and $\ (n,M^{2})-$planes we have
different series or trajectories in $(L,M,n)-$space for $n^{2S+1}L_{J}-$
meson states.

There is a $(\pi \rho )-$ series for $n^{3}L_{L-1}-$meson states with $P=(-1)^{L+1}$%
, $C=(-1)^{L+1}$. Mass values of the members of the $(\pi \rho )-$ series
are determined by the formula:
\begin{equation}
M_{n,L}^{\pi \rho}=c+\varkappa \sqrt{2(n^{r}+L)-1}+\varkappa \sqrt{2(n^{r}+L)}
\end{equation}

There is a $\pi -$ series for $n^{1}L_{L}-$meson states with $P=(-1)^{L+1}$, $%
C=(-1)^{L}$. Mass values of the members of the $\pi -$ series are determined
by the formula:
\begin{equation}
M_{n,L}^{\pi}=2c+2\varkappa \sqrt{2(n^{r}+L)}
\end{equation}

There is a $(\rho \pi )-$ series for $n^{3}L_{L}-$meson states with $P=(-1)^{L+1}$, 
$C=(-1)^{L+1}$. Mass values of the members of the $(\rho \pi )-$ series are
determined by the formula:
\begin{equation}
M_{n,L}^{\rho\pi}=c+\varkappa \sqrt{2(n^{r}+L)}+\varkappa \sqrt{2(n^{r}+L)+1}
\end{equation}

There is a $\rho -$ series for $n^{3}L_{L+1}-$meson states with $P=(-1)^{L+1}$, $%
C=(-1)^{L+1}$. Mass values of the members of $\rho -$ series are determined
by the formula:
\begin{equation}
M_{n,L}^{\rho}=2\varkappa \sqrt{2(n^{r}+L)+1}
\end{equation}

With the help of these formulae for the series presented a few mass
relations, which are not dependent on intrinsic quantum numbers $L$ and $n$
easily can be obtained. For instance, mass values of members of $\pi -$
series, $(\rho \pi )-$ series and $\rho -$ series with the same $L$ obey the
following mass relation:
\begin{equation}
2M_{n,L}^{\rho\pi}(n^{3}L_{L})=M_{n,L}^{\pi}(n^{1}L_{L})+M_{n,L}^{\rho}(n^{3}L_{L+1})
\label{mnl}
\end{equation}

It is interesting, that a quantity $\tau $ for members of $P$-wave meson
multiplets, which played an important role in a determination of 
properties of spin-dependent forces between heavy quarks and antiquarks \cite%
{diek,lucha}, have the following value in the case considered (remind that
we neglect the difference between masses of $u-$ and $d-$quarks).
\begin{equation}
\tau =\frac{M_{n,L}^{\rho}(n^{3}L_{L+1})-M_{n,L}^{\rho\pi}(n^{3}L_{L})}{%
M_{n,L}^{\rho\pi}(n^{3}L_{L})-M_{n,L}^{\pi}(n^{3}L_{L-1})}=\frac{\varkappa \sqrt{2(n^{r}+L)+1%
}-c-\varkappa \sqrt{2(n^{r}+L)}}{\varkappa \sqrt{2(n^{r}+L)+1}-\varkappa 
\sqrt{2(n^{r}+L)-1}}
\end{equation}

So for members of $P-$wave superlight meson multiplets with $n^{r}=0$ $\tau $ is
equal to\ $0.19$. For the $c\overline{c}$ - states in $P$ $-$wave the value $%
\tau $ is $\sim 0.5$, while for the $b\overline{b}$ - states it is 
$\sim 0.65$.

\medskip 
\textbf{3. Evaluation of meson masses and comparison with experimental data}
\smallskip

When evaluating masses of unknown excited meson states, we use the formulae
written above together with the values of two parameters $c$ and $\varkappa $%
, which have been obtained by fitting of mass values of experimentally
detected meson states. The $c$ and $\varkappa $ values obtained 
by this manner are $c=69$ $%
MeV,$ $\varkappa =382\pm 4$ $MeV$. Then a mass value of an ordinary
superlight $\overline{q}q^{\prime }-$meson, which consist of  $u-$, $d-$%
quarks and antiquarks and have the isotopical spin equal to unity, can be
evaluated with an absolute uncertainty less than $40MeV$.

The results obtained with $c=69$ $MeV,$ $\varkappa =385$ $MeV$ are shown in
the Tables 1 and 2, where we present the mass values of orbital excitations
of superlight meson states up to $L=4$ and the radial excitations of 
superlight pseudoscalar,
vector and scalar meson states up to $n^{r}=4$. Although we restricts us with
those meson states, which have isotopical spin values equal to unity, in
order to exclude the unknown influence of mixing with superfluous states in $%
I=0$ sector, nevertheless it is found that an account of mixing in $I=1$ sector
is also necessary. As it follows from the Tables 1 and 2, the masses
evaluated for first radial and orbital excitations of the vector $1^{--}$
mesons have no reliable confirmations by existing data. May be it is due to
the shortcoming of the mass formulae in this region. It seems rather
unlikely because of other evaluated mass values especially for the orbital 
excitations coincide with the data \cite{pdg} within the experimental
errors. Thus we assume that mixing between the first radial and orbital
excitations for the $1^{--}$ mesons take place. However, it is possible,
that the more complicated situation should be considered, when in the mass
region $\sim $ $1.5$ $GeV$ mixing between standard $1^{--}$ mesons and
vector non $\overline{q}q^{\prime }-$mesons take place as well. Below we
consider only mixing between the first radial and orbital excitated
standard  $1^{--}$ mesons. In this case the experimentally visible vector meson
states $V_{R}^{\prime }$ and $V_{O}^{\prime }$ are connected with the pure
ones through a rotation on some angle $\theta $:
\begin{equation}
\begin{array}{l}
V_{R}^{\prime }=V_{R}\cos \theta -V_{O}\sin \theta , \\ 
V_{O}^{\prime }=V_{R}\sin \theta +V_{O}\cos \theta%
\end{array}%
\end{equation}

Taking into account the existing experimental data the following cases are
most preferable. In the first case the experimentally observed bump at $%
1465MeV$ consists of one resonance $V_{R}^{\prime }$, that is mainly the
radial excitation of $\rho -$meson. In the second case the experimentally
observed bump at $1465MeV$ consist of two resonances $V_{R}^{\prime }$ and $%
V_{O}^{\prime }$. Because of we know the masses of $V_{R}$ and $V_{O}$ (see
Tables 1 and 2), mixing angles and mass values of $V_{R}^{\prime }$ and $%
V_{O}^{\prime }$ in these two cases easily can be evaluated. In the first
case the $V_{O}^{\prime }$ resonanse lie below the $V_{R}^{\prime }$ and
have mass $\sim 1350$ $MeV$, while in the second case the $V_{R}^{\prime }$
and $V_{O}^{\prime }$ have approximately equal masses $\sim 1450$ $MeV$. 
Take into account this consideration we expect that in the mass region
between \ 1.3 GeV and 1.6 GeV two standard $\overline{q}q^{\prime }-$meson
resonances with $J^{PC}=1^{--}$ and $I=1$ exist, namely the first radial and
orbital excitations mixing each another. The similar conclusion was
presented in the paper \cite{c4}.  In the general case,
as mentioned above, the most complicated situation may arise, which is not 
disscussed here, when the mixing with cryptoexotic states can occur. 
This case demands further investigation.

Let us compare with the data \cite{pdg} the mass relation (\ref{mnl}), which must
obey for the mesons with $L\neq 0$ and $S=1$. At present this relation can
be checked only for the mesons with $L=1.$ After the substitution the
experimental values of the masses for the $a_{1}(1260)$, $b_{1}(1235)$ and $%
a_{2}(1320)$ mesons, we obtain, that this relation is fulfilled with account
of experimental uncertanties. Moreover, the mass formulae presented above 
permit to explain  degeneracy with respect to mass values for different 
$J^{PC}$ mesons. For example, the mass formulae  give 
$M(\rho^{\prime\prime})=M(\rho_3)$. If one use 
the data, then  $1720\pm20 MeV=1688.8\pm2.1 MeV$. In the high mass value region
the degeneracy of such kind will be increased, as it is seen in the $\sim 2315 MeV$
region, where the $0^{-+}$, $1^{--}$, $4^{-+}$, $4^{--}$ and $5^{--}$ mesons 
  are predicted.

In the framework of nonrelativistic potential models the following relation
for the mass values of the members of $P$ multiplets is well known
\begin{equation}
M(^{1}P_{1})-5/9M(^{3}P_{2})-3/9M(^{3}P_{1})-1/9M(^{3}P_{0})=0  \label{pform}
\end{equation}

Using the mass formulae for the superlight mesons with $n^{r}=0$ written
above, this relation goes into the form:
\begin{equation}
14(c+\varkappa \sqrt{2})MeV=\varkappa (13\sqrt{3}+1)MeV
\end{equation}

When substituting the numerical $c$ and $\varkappa $ values, we obtain $%
8589 MeV=9054 MeV$. Thus, the relative error for the fulfillment
of this relation for superlight
mesons is $\sim 5\%$. This result has been presented for illustration only
the numerical resemblance between the  potential approach and the 
phenomenological one. Notwithstanding an
 approximated character of the relations (15) and (16)  one can see,
that these two approaches do not contradict each another. The
comparison of the results presented in the Table 2 for the mass values of
the radial excitations with the data, so as with the evaluations of the mass
values of the radial excitations for the vector and pseudoscalar mesons with
the help of Dirac equation with the potential $\sim -a/r+\gamma ^{0}kr$ \cite%
{c3,c5}, also confirms the previous point. Moreover the obvious fact should be 
noted, that the accuracy of evaluations 
 with the help of formulae (\ref{mnf}), (\ref{enf}) and (\ref%
{jnf}) for the radial excitated mesons are worse that for orbital ones.
For this reason we use in Table 2 the maximal systematical uncertainty
for mass values.

\medskip 
\textbf{4. Discussion and conclusions} \smallskip

It is useful to compare the mass values evaluated with the phenomenological
mass formulae  (\ref{mnf}), (\ref{enf}) and (\ref%
{jnf})   and shown in the Tables 1 and 2 with the values obtained
in the framework of other approaches. As it was noted in the previous
section the evaluated values are in a concordance with the 
values obtained in the relativistic model of
quasi-independent quarks moving in the potential $\sim -a/r+\gamma ^{0}kr$ 
\cite{c3,c5}. However, the mass values of the first radial of $\rho $ - 
meson and orbitally
excitated vector meson  in the framework of  the phenomenological
approach considered show a discrepancy with data. We offer the mixing of these
states in order to put mass values into
accordance with data. It is possible,  that at present the experimental data 
and their interpretation are not
complete in this region and additional efforts are needed to clarify the
situation. For instance, in the framework of the well-known potential
model \cite{godf}
the mass values of $2$ $^{3}S_{1}$, $1^{3}D_{1}$ and $3$ $^{3}S_{1}$ states
lie considerably higher (at $1.45$,$1.66$, $2.00$ $GeV$ correspondingly)
than the predictions of the method considered. Moreover, in Ref. \cite{faust}
the mass value $1486 MeV$ has been evaluated  the first radial 
excitation of $\rho $ - meson.

Another importain problem is the interpretation of mesons discovered in
scalar channel. Although, the mass values of the $a_{0}(980)$ and\ $\
f_{0}(980)$ mesons show the ideal mixing between them, the problems with the
intensities of different decay modes exist (in particular the$\ K\overline{K}
$ decay mode enhancement). The most widespread explanation of these
facts now consist in
the four-quark nature of the $a_{0}(980)$ and\ $\ f_{0}(980)$ mesons \cite%
{jaffe,acha,diak}. However, if one use  typical mass values for scalar
diquarks \cite{berg},  estimations of  mass values for lowest
diquark-antidiquark scalar mesons give $\sim 1200\div 1400MeV$. The results
of our evaluations (Table 1) also support the existence of the $P-$wave 
\textbf{\ }$\overline{q}q^{\prime }-$mesons with masses $\sim 980$ $MeV$.
There is an accidental degeneracy of their mass values with the value of $%
\overline{K}K-$threshold, which complicates a mechanism of an
extraction of $a_{0}(980)$ and\ $\ f_{0}(980)$ decays characteristics \cite%
{baru}. Moreover it is possible, that  mixing of these mesons with a $%
\overline{K}K-$molecule arises \cite{kal}. The complementary reasoning in
favour of the existence of the $P-$wave \textbf{\ }$\overline{q}q^{\prime }-$%
mesons with masses $\sim 980$ $MeV$ is the coincidence of the evaluated mass value
for the first radial excitation of $a_{0}(980)$ meson (Table 2) with the
mass value of the $a_{0}(1450)$ meson \cite{pdg}.

The problems discussed above are the problems of the phenomenological scheme
for meson mass evaluations
considered, as well as they represent the actual and unsolved during a long
time problems of light meson spectroscopy.

\smallskip\ \noindent \textbf{Acknowledgments}. The author is grateful to
Yu.V.~Gaponov, V.I.~Savrin and S.V.~Semenov for useful discussions.


\newpage \bigskip {\small Table 1. }

{\small 
}

{\small 
\parbox[t]{16.cm}{Evaluated masses in MeV for the ground states  and the
orbital excitations of the   $\bar qq'-$ mesons in
comparison with the  data from Ref.\cite{pdg}. }}

\medskip

\begin{tabular}{cccccccc}
\hline
Meson & $J^{PC}$ & $M^{exp}(MeV)^{[17]}$ & $M^{ev}(MeV)$ & Meson & $J^{PC}$ & 
$M^{exp}(MeV)^{[17]}$ & $M^{ev}(MeV)$ \\ \hline
$\pi $ & 0$^{-+}$ & 138$\pm $3.1 & 138$\pm $30 & $\rho _{3}$ & 3$^{--}$ & 
1688.8$\pm $2.1 & 1722$\pm $30 \\ 
$\rho $ & 1$^{--}$ & 775.8$\pm $0.5 & 770$\pm $30 & $a_{2}^{3}$ & 2$^{++}$ & 
- & 1873$\pm $30 \\ 
$a_{0}$ & 0$^{++}$ & 984.7$\pm $1.2 & 998$\pm $30 & $b_{3}^{3}$ & 3$^{+-}$ & 
- & 2024$\pm $30 \\ 
$b_{1}$ & 1$^{+-}$ & 1229.5$\pm $3.2 & 1227$\pm $30 & $a_{3}^{3}$ & 3$^{++}$
& - & 2031$\pm $30 \\ 
$a_{1}$ & 1$^{++}$ & 1230$\pm $40 & 1280$\pm $30 & $a_{4}$ & 4$^{++}$ & 2040$%
\pm $12 & 2037$\pm $30 \\ 
$a_{2}$ & 2$^{++}$ & 1318.3$\pm $0.6 & 1334$\pm $30 & $a_{3}^{4}$ & 3$^{--}$
& - & 2177$\pm $30 \\ 
$a_{1}^{2}$ & 1$^{--}$ & - & 1506$\pm $30 & $b_{4}^{4}$ & 4$^{-+}$ & - & 2316%
$\pm $30 \\ 
$\pi _{2}$ & 2$^{-+}$ & 1672.4$\pm $3.2 & 1678$\pm $30 & $a_{4}^{4}$ & 4$%
^{--}$ & - & 2313$\pm $30 \\ 
$a_{2}^{2}$ & 2$^{--}$ & - & 1700$\pm $30 & $a_{5}^{4}$ & 5$^{--}$ & - & 2310%
$\pm $30 \\ \hline
\end{tabular}

\bigskip

\bigskip

\smallskip

{\small Table 2. }

{\small 
}

{\small 
\parbox[t]{16.cm}{Evaluated masses in MeV for  the
radial excitations (from $n^r=1$ to $n^r=4$) of the pseudoscalar, vector and scalar  $\bar qq'-$ mesons in
comparison with the  data from Ref.\cite{pdg}. }}

\medskip

\begin{tabular}{cccccccc}
\hline
Meson & $J^{PC}$ & $M^{exp}(MeV)^{[17]}$ & $M^{ev}(MeV)$ & Meson & $J^{PC}$ & 
$M^{exp}(MeV)^{[17]}$ & $M^{ev}(MeV)$ \\ \hline
$\pi ^{\prime }$ & 0$^{-+}$ & 1300$\pm $100 & 1227$\pm $40 & $\rho ^{\prime
\prime \prime }$ & 1$^{--}$ & - & 2037$\pm $40 \\ 
$\pi ^{\prime \prime }$ & 0$^{-+}$ & 1812$\pm $14 & 1678$\pm $40 & $\rho
^{IV}$ & 1$^{--}$ & - & 2310$\pm $40 \\ 
$\pi ^{\prime \prime \prime }$ & 0$^{-+}$ & - & 2024$\pm $40 & $%
a_{0}^{\prime }$ & 0$^{++}$ & 1474$\pm $12 & 1506$\pm $40 \\ 
$\pi ^{IV}$ & 0$^{-+}$ & - & 2316$\pm $40 & $a_{0}^{\prime \prime }$ & 0$%
^{++}$ & - & 1873$\pm $40 \\ 
$\rho ^{\prime }$ & 1$^{--}$ & 1465$\pm $25 & 1334$\pm $40 & $a_{0}^{\prime
\prime \prime }$ & 0$^{++}$ & - & 2177$\pm $40 \\ 
$\rho ^{\prime \prime }$ & 1$^{--}$ & 1720$\pm $20 & 1722$\pm $40 & $%
a_{0}^{IV}$ & 0$^{++}$ & - & 2441$\pm $40 \\ \hline
\end{tabular}

\bigskip

\bigskip

\end{document}